# Intelligent Meta-Imagers: From Compressed to Learned Sensing

Chloé Saigre-Tardif[1,2,*], Rashid Faqiri[1,3,*], Hanting Zhao[4], Lianlin Li[4], and Philipp del Hougne[1,†]

[1] Univ Rennes, CNRS, IETR - UMR 6164, F-35000, Rennes, France

[2] INSA Rennes, CNRS, IETR - UMR 6164, F-35000, Rennes, France

[3] The Blackett Laboratory, Department of Physics, Imperial College London, London SW7 2AZ, UK

[4] State Key Laboratory of Advanced Optical Communication Systems and Networks, Department of Electronics, Peking University, 100871 Beijing, China.

[*] These authors contributed equally.

[†] Correspondence to philipp.del-hougne@univ-rennes1.fr.

## Abstract

Computational meta-imagers synergize metamaterial hardware with advanced signal processing approaches such as compressed sensing. Recent advances in artificial intelligence (AI) are gradually reshaping the landscape of meta-imaging. Most recent works use AI for data analysis, but some also use it to program the physical meta-hardware. The role of "intelligence" in the measurement process and its implications for critical metrics like latency are often not immediately clear. Here, we comprehensively review the evolution of computational meta-imaging from the earliest frequency-diverse compressive systems to modern programmable intelligent meta-imagers. We introduce a clear taxonomy in terms of the flow of task-relevant information that has direct links to information theory: compressive meta-imagers indiscriminately acquire all scene information in a task-agnostic measurement process that aims at a near-isometric embedding; intelligent meta-imagers highlight task-relevant information in a task-aware measurement process that is purposefully non-isometric. The measurement process of intelligent meta-imagers is thus simultaneously an analog wave processor that implements a first task-specific inference step "over-the-air". We provide explicit design tutorials for the integration of programmable meta-atoms as trainable physical weights into an intelligent end-to-end sensing pipeline. This merging of the physical world of metamaterial engineering and the digital world of AI enables the remarkable latency gains of intelligent meta-imagers. We further outline emerging opportunities for cognitive meta-imagers with reverberation-enhanced resolution and we point out how the meta-imaging community can reap recent advances in the vibrant field of metamaterial wave processors to reach the holy grail of low-energy ultra-fast all-analog intelligent meta-sensors.

# CONTENTS

# 1. Introduction

Electromagnetic (EM) computational meta-imagers are poised to become the backbone of emerging cyber-physical systems that drive important societal developments in security, autonomous mobility, smart health care and touchless human-machine interaction. Modern meta-imagers combine the virtue of low-cost metamaterial hardware with the speed and agility needed for task-specific *in situ* instantaneous sensing and monitoring. Their programmability has been the key to drastic latency improvements by enabling task-aware "intelligent" measurements that pre-select useful information. In this Review, we retrace the evolution of computational meta-imagers from early frequency-diverse compressive-sensing approaches to state-of-the-art programmable learned-sensing techniques. By following the flow of information through the sensing pipeline, we explain this evolution in terms of the elimination of redundant and irrelevant information at the earliest possible stage: the measurement. These developments evidence the power of merging the physical world of metamaterial engineering with the digital world of artificial intelligence (AI), and we conclude by looking forward toward all-analog cognitive meta-imagers with unprecedented resolution.

Our modern society increasingly develops needs for remote probing of people's location, arming, activity, body language and even their physiological state. Ideally, this intelligent surveillance can be realized (i) remotely, (ii) without requiring the subjects to cooperate (e.g., by carrying tags), (iii) without dependence on the subjects' skin color or clothing, (iv) without dependence on scene lighting or fog, (v) without ionizing radiation, and (vi) with minimal visual privacy infringements. EM waves are predestined to fulfill these requirements. Some progress toward wireless surveillance has been achieved by exploiting the channel state information of existing WiFi infrastructure[1–5], but in general the degrees of freedom of such setups, and hence their ability to probe the scene, are quite limited. Therefore, most wireless surveillance systems rely on antenna arrays, often in combination with wideband signals[6–14]. However, such archetypical implementations of the required EM sensor aperture are costly in the case of antenna arrays[15] or slow in the case of mechanically scanned synthetic apertures[16].

Metamaterial hardware has enabled alternative low-cost aperture architectures, referred to as "meta-imagers". These coded metamaterial apertures multiplex scene information across diverse measurement modes, offered by the metamaterial's degrees of freedom, onto a single (or few) detector(s); however, these meta-imagers rely on computationally highly expensive inverse-scattering solvers to reconstruct a scene image. To expedite the entire sensing process, on the one hand, the deployed methodologies for data analysis have been refined. In particular, the realization that the ultimate goal is almost always *not* a visual representation of the scene but an answer to a specific task has led to the reconstruction-free interpretation of raw measured data. On the other hand, the emergence of programmable metamaterials has recently enabled the sculpting of intelligent scene illumination modes that select, hand-in-hand with the subsequent data processing algorithm, task-relevant information. Thereby, both the number of measurements as well as the digital processing burden are simultaneously minimized, yielding a double advantage in terms of latency as well as improvements on other relevant metrics like data storage and power consumption. State-of-the-art EM meta-imagers now enable intelligent, high-fidelity, low-cost and real-time sensing and monitoring that can be adapted *in situ* to task and scene – in line with the requirements of the targeted applications in security, autonomous mobility, smart health care and touchless human-machine interaction.

In this Review, we present a comprehensive and easy-to-follow taxonomy of influential meta-imagers that focusses on the flow of information through the sensing pipeline and offers a clear criterion to distinguish between compressive and intelligent meta-imagers. None of the four review papers[17–20] related to meta-imaging in the recent literature offer these insights; in fact, they do not cover intelligent meta-imagers in which AI defines the *physical* configuration of the programmable meta-imager at all. Section 2 of our paper is dedicated to compressive meta-imagers which aim at an indiscriminatory near-isometric embedding of all scene information in a latent space. After covering the fundamentals of computational imaging modalities, we classify compressive meta-imagers in terms of the utilized scene illumination principle and the deployed metamaterial hardware. Section 3 introduces the modern twist of meta-imaging based on learned sensing: intelligent meta-imagers that discriminate between task-relevant and task-irrelevant information in a purposefully non-isometric measurement process. We explain different approaches for the integration of programmable meta-atoms as physical weights into hybrid analog-digital end-to-end learned-sensing pipelines, and we benchmark the performance of such intelligent meta-imagers in terms of latency against the conventional compressive scene illumination paradigms from section 2. Finally, in section 4, we lay out future avenues for research on intelligent meta-imagers. We discuss (i) how futuristic meta-imagers may be endowed with cognitive features, (ii) how they can leverage indoor reverberation for unprecedented resolution, and (iii) how conceptual links to the field of analog computing with metamaterials can be pushed toward achieving ultra-fast low-power all-analog implementations of meta-imagers on easily integrable metamaterial platforms.

## 2. Compressive Meta-Imagers

**2.1 Fundamentals.** Imaging and sensing can be broadly defined as the characterization of a scene based on how it scatters waves. The important steps of a conventional generic sensing pipeline are outlined in Figure 1. During the measurement process, the scene $S$ is illuminated with waves and some of the scattered waves are measured, yielding the measurement data $D$. The latter is then used to reconstruct an image of the scene, which is finally interpreted to answer the specific task that motivated the entire sensing process. We will return to the importance of realizing that usually the entire sensing process is conducted to answer a specific task, rather than to obtain a high-fidelity visual image of the scene for human consumption, below. To start, however, we take a closer look at the measurement process. The scene is usually assumed to be in free space such that all scattering of the incident waves is due to the scene, and, moreover, multiple scattering within the scene is typically neglected. These assumptions are known as the first Born approximation, and imply that the measurement process is a linear mapping of the (vectorized) scene $S$ via a sensing matrix $\mathbf{H}$ into the measured data vector $D$:

$$D = \mathbf{H}S + N, \qquad (1)$$

where the vector $N$ denotes detector noise[21]. The sensing matrix $\mathbf{H}$ contains the (vectorized) measurement modes that probe the scene and multiplex the scene information onto the detector(s). Throughout this paper, we use the term "scene illuminations" to refer to these measurement modes which are the dot product of the (vectorized) transmitted and received electric fields at the scene voxels[21]. Meta-imagers deploy metamaterial hardware on the transmitting and/or receiving side, and due to reciprocity, no distinction between metamaterial transmitter and metamaterial receiver is needed.

The nature of the sensing matrix $\mathbf{H}$ depends on the choice of scene illumination patterns. Conventional imagers establish a one-to-one mapping between scene voxels and image pixels, such that the structure of $\mathbf{H}$ is often as simple as that of an identity matrix, implying that image reconstruction from the measured data is trivial. In these conventional approaches, the scene is scanned with diffraction-limited spots, as illustrated in Figure 1. Important contributions from the metamaterial community in the realm of conventional imaging include ideas to enhance the image resolution, usually by capturing evanescent waves, such as the perfect lens[22], the super-lens[23], or the hyper-lens[24]; these developments are outside the scope of the present Review since they involve manipulations in close proximity of the scene – which is not compatible with the requirement for non-invasive remote sensing and monitoring of the EM imaging applications we cover here. Interested readers may refer to a recent review in Ref.[18] for more details.

Computational imaging[25], in contrast to conventional imaging, multiplexes the scene across different masks of a coded aperture onto a single (or few) detector(s)[26,27], as illustrated in Figure 1. In the simplest implementation, each mask has a random pattern.

Each measurement thus contains contributions from all scene voxels, weighted by the utilized mask, such that the sensing matrix **H** is highly complicated. Such computational "single-pixel cameras"[26,28] alleviate the very high hardware cost of the dense reconfigurable antenna arrays required to implement an EM scannable focal spot in conventional EM imaging. The signal-to-noise ratio (SNR) of multiplexed measurements is also advantageous because reflections from a large portion of the scene (as opposed to a single voxel) are collected in each measurement. However, "single-pixel cameras" face the computational challenge of reconstructing the scene based on the multiplexed measurements.

Most natural scenes are sparse, either directly in real space or in a transform domain, or in terms of their gradients[29], or in blocks[30], or in other representations[31,32]. Compressive imagers, a subset of computational imagers, take advantage of this inherent sparsity and faithfully reconstruct the scene with an underdetermined sensing matrix (using fewer measurements than scene voxels)[33]. Wave propagation during the measurement process acts as an analog (physical) compressor of the scene information. Equation 1 describes this compression as a linear embedding of the probed scene in a lower-dimensional space. Ideally, the compression process does not distort the relative distances between any two points such that the loss of information due to the compression is minimal. Formally, this idea of a distance-preserving transformation is known as isometry in signal processing theory where a central result is the restricted isometry property[34,35]: if the sensing matrix satisfies this property for a given level of input signal sparsity, then such inputs can be reconstructed stably despite the sensing matrix being underdetermined. The spirit of this theorem is that even though the sensing matrix may not be isometric, it can behave as if it was nearly isometric when acting on sparse vectors. In particular, random matrices have been shown to satisfy this property[36], at least when the problem dimensions are sufficiently high. This result is of high importance to identify physical platforms capable of implementing suitable sensing matrices.

Along these lines, the relevance of metamaterial engineering to the field of EM computational imaging emerges since 2013[37]: the degrees of freedom offered by metamaterial hardware can be conveniently leveraged to implement the required coded-aperture masks. We provide a comprehensive overview of the diverse proposals of compressive meta-imagers in the following subsection. We carefully sort these works in terms of the utilized scene illumination strategy (which is the key factor in terms of the flow of information through the sensing pipeline) and the underlying metamaterial hardware class. Initially, computational meta-imagers relied on spectral diversity; with the advent of programmable metamaterials, they increasingly use configurational diversity instead. Toward the end of the next subsection, we will see how the metamaterial's programmability can be leveraged to synthesize scene illumination patterns. Subsection 2.2 partially overlaps with Refs.[18,19] that recently reviewed compressive meta-imagers, but Refs.[18,19] cover neither our taxonomy in terms of scene illumination strategy nor in its entirety the breadth of metamaterial hardware implementations that we discuss.

This synthesis ability prepares the ground for the subsequent section on intelligent meta-imagers. Hardware programmability is ultimately the basis for merging the physical world of metamaterial engineering with the digital world of AI in the context of computational imaging. The crucial difference between compressive and intelligent meta-imagers lies in how the programmability is used. While compressive meta-imagers seek a sensing matrix that acts as isometric projection of the probed scene in order to faithfully compress *all* scene information, an intelligent meta-imager seeks a non-isometric projection that compresses and *simultaneously distorts in a purposeful manner* to facilitate answering the specific task. The physical layer thus performs "intelligent" task-specific analog signal processing in the latter but not in the former case. Note that in this Review we reserve the attribute "intelligent" for meta-imagers in which the AI impacts the physical measurement process which is not the case if AI intervenes purely for post-measurement data analysis.

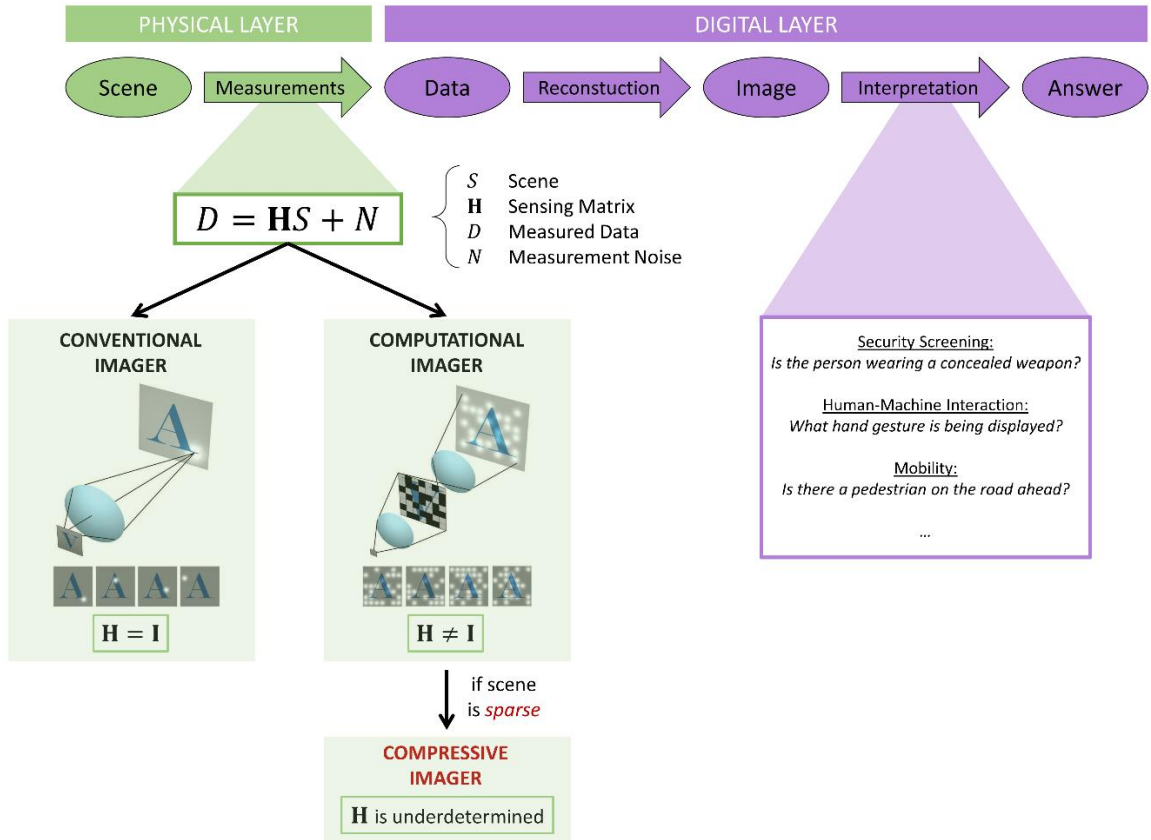

**Figure 1. Conceptual basis of a conventional, computational and compressive imager.** The generic sensing pipeline consists of three steps. First, information about the scene $S$ is encoded on the physical layer via measurements in data $D$. Under the common Born approximation, this is a linear mapping via the sensing matrix $\mathbf{H}$. Second, an image of the scene is reconstructed from the data on the digital layer. Third, the image is interpreted to answer the specific task that motivated the entire sensing process. A conventional imager uses diffraction-limited spots as measurement modes, such that the reconstruction step is trivial because $\mathbf{H}$ is an identity matrix. In contrast, a computational imager uses diverse scene illumination patterns as measurement modes, resulting in a non-trivial sensing matrix and image reconstruction. If the scene is sparse, image reconstruction can be achieved with an underdetermined sensing matrix, known as compressive imaging. The bottom left schematics are adapted with permission from *Science* **339**, 310 (2013).[37] Copyright 2013 American Association for the Advancement of Science.

**2.2 Implementations.** Our taxonomy of compressive meta-imagers begins by asking a high-level question: what is the nature of the resulting scene illumination patterns? The most common case is that they are pseudo-random. We use the term "pseudo-random" instead of "random" on purpose because in most physical realization of random patterns some deterministic component that is not randomly stirred in different realizations remains, such that the patterns do not follow ideal random distributions (see, for instance, Refs.[38,39]). A first class of pseudo-random patterns leverages spectral diversity: by sweeping the frequency, the scene illumination patterns can drastically vary but the scene reflectivity remains roughly constant. This broadband operation implies substantial hardware and spectrum allotment cost, which is alleviated in the second class of pseudo-random patterns based on configurational diversity. In this second class, the scene is illuminated at a single frequency with different patterns by electronically reconfiguring some scattering elements of the aperture. Either way, pseudo-random patterns have by definition a finite overlap, even if they were perfectly random, such that subsequent measurements inevitably acquire to some extent redundant information. The acquisition of redundant information can be prevented by using orthogonal scene illumination patterns. This second category of scene illuminations can be implemented if the hardware is programmable. Orthogonal patterns can be based on any basis other than the canonical one because in the canonical basis ($\mathbf{H}=\mathbf{I}$) the imager is not compressive. A well-known orthogonal basis is the Hadamard basis, but orthogonal patterns can also use any other basis. Specifically, the latter implies that upon visual inspection, orthogonal patterns cannot always be distinguished from pseudo-random patterns. Their orthogonality can be confirmed by computing the singular value decomposition of the sensing matrix, which is perfectly flat as opposed to downward sloping for orthogonal as opposed to pseudo-random patterns. Alternatively, the overlap of any pair of scene illumination patterns can be computed and should be zero as opposed to finite. The third category is a special case of orthogonal patterns for which the base is that of principle scene components. The latter can be obtained via a principle component analysis (PCA)[40] of a library of representative scenes. Unlike pseudo-random or orthogonal patterns, principle scene components are thus specific to the scene and not generic. Principle scene components are optimally efficient for a near-isometric embedding – and hence by definition agnostic to the specific task.

Before diving into metamaterial implementations of these scene illumination strategies, we summarize in the first row of Figure 2 representative conventional implementations of these ideas for compressive scene illuminations. Conventionally, pseudo-random patterns are implemented with configurational diversity, as seen in Figure 2a for a single-pixel terahertz camera[41]. In the optical regime where detector arrays (e.g. CCDs) are standard inexpensive equipment, compressive imagers leveraging spatial diversity have also been reported: light reflected from a scene can be multiplexed across a random scattering medium onto an array of detectors[42]. Using an array of individually controllable antennas, principle scene components can be synthesized, a numerical study of which is seen in Figure 2b[43].

*2.2.1 Frequency-Diverse Pseudo-Random Patterns.* − The very first meta-imager is seen in Figure 2c[37]. It consists of a 1D waveguide that is patterned with metamaterial elements (referred to as meta-atoms). The latter are resonant cELC elements[44,45] such that the guided wave couples to the meta-atoms and thereby leaks out of the waveguide into the far field, where it illuminates the scene. In order to illuminate the scene with a random (speckle-like) pattern that varies as a function of frequency (see Figure 2c), the geometrical properties (and thereby the resonances) of the meta-atoms differ along the 1D waveguide. This leaky 1D waveguide constitutes a frequency-diverse meta-imager capable of generating pseudo-random illumination patterns. The amount of reasonably different pseudo-random scene illuminations that can be accessed within a given frequency interval depends on the sharpness of the meta-atoms' resonances. However, the latter cannot be increased arbitrarily, in particular because such an increase is accompanied by a decrease in energy radiated into the scene.

Considerable improvements in the frequency diversity were subsequently unlocked by moving from a 1D leaky waveguide to an overmoded 3D leaky chaotic cavity[46] (see Figure 2e). The latter supports a high density of modes, and a perforated cavity surface allows waves to leak out into the scene. Because chaotic cavities completely mix polarizations, at a given frequency the scene illumination's various polarization components are very different, a feature that enables polarimetric computational meta-imaging[47]. 3D leaky chaotic cavities have moreover been used for computational imaging of (primary) sources as opposed to reflecting objects (secondary source): in Ref.[48] to image thermal sources, and in Ref.[49] for direction of arrival (DoA) estimation. The improved diversity of 3D leaky chaotic cavities comes at the price of a bulkier device, however, and later a compromise between the two previously discussed options was found: a quasi-2D leaky chaotic cavity. The latter is shown in Figure 2h and is electrically large in two dimensions whereas its third dimension is much smaller than the wavelength (hence "quasi-2D"); one surface contains sub-wavelength slots through which waves leak out into the scene. Based on this compact meta-imager hardware, a remarkable system-level implementation capable of imaging human-sized scenes was developed in Ref.[50]. Successful operation of this multistatic system involving 24 transmitting and 72 receiving frequency-diverse 2D leaky chaotic cavities requires elaborate careful calibration procedures, as detailed in Ref.[50]. Instead of sequentially switching through all possible transmitter-receiver combinations, the system can also be operated in a massive-MIMO mode with simultaneous, mutually orthogonal coded transmit signals[51]. The hardware simplicity of leaky 3D and 2D cavities is attractive especially for operation at higher frequencies (50-65 GHz in Ref.[52] and 75-110 GHz in Ref.[53]) where implementing the configurational diversity discussed in subsequent sections is more challenging than at microwave frequencies.

So far, we have seen three examples in which the feed was inside the frequency-diverse metamaterial structure. Another approach, seen in Figure 2j, is to use an ultrathin metasurface that transmits/reflects a wave generated by a distant horn antenna to the scene. An obvious inconvenience here, in contrast to the previously outlined techniques, appears to be the lack of compactness if metasurface, horn antenna, and the distance between the two are considered. In Ref.[54] this distance was in fact desired because the metasurface was placed in close proximity to the scene in order to couple evanescent waves to the far field. A similar approach has also been implemented in the optical regime[55]. In the ultrasound regime, a metasurface transmit-array was used as frequency-diverse coded aperture for compressive imaging[56]. We will see later-on with respect to our interest in remote EM sensing and monitoring that the compactness concern vanishes when stray ambient waves (e.g., WiFi) are recycled by the metasurface for EM imaging.

As last example of frequency-diverse meta-imagers, we look in Figure 2m at disordered bulk collections of resonators, sometimes referred to as "resonant metalenses"[57–59]. This technique has been mainly explored in acoustics to place the receiver inside such a frequency-diverse system for multi-speaker listening[57,60], or to place the system very close to the scene to couple evanescent waves to the far-field[58]. The concept has also been extended to elastic vibrations[61].

*2.2.2 Configuration-Diverse Pseudo-Random Patterns.* – Now, instead of relying on different frequencies yielding distinct scattering responses, we consider meta-imagers whose scattering responses can be electronically reconfigured, thereby enabling access to a series of pseudo-random scene illuminations despite single-frequency operation. The latter is advantageous in terms of hardware cost and spectrum allotment considerations. The concept of programmable metamaterials[62–65] enabled this milestone in meta-imager hardware. Indeed, programmability was the key hardware improvement and all subsequent improvements on meta-imagers that we discuss in the next subsections and sections mainly relate to reaping the benefits of this programmability as "intelligently" as possible. Note that single-frequency operation does not necessarily result in a complete loss of range resolution[19]: the well-known link between bandwidth and range resolution[66] is only applicable in the absence of obliquely incident illumination. Many compressive meta-imagers operate at distances from the scene that are small enough such that waves are also incident at oblique angles[67], and sometimes the meta-imager is even moved which improves the range resolution further[68]. Once we go beyond the assumption of a scene surrounded by free space under which we are operating thus far, we will see in subsection 4.2 that even deeply sub-wavelength resolution with single-frequency measurements is possible inside complex scattering enclosures[69] – without any access to evanescent waves.

The first configuration-diverse programmable meta-imager[70] emerged only two years after the first frequency-diverse meta-imager[37], and similarly relies on a 1D leaky waveguide, as seen in Figure 2d. Here, each meta-atom is loaded with an individually addressable PIN-diode. Leakage of the guided wave via the meta-atom into the far field can be switched on or off by controlling the bias voltage of the PIN diode. Different "coding patterns" of the programmable meta-atoms thereby realize distinct pseudo-random scene illumination patterns at a single frequency, as seen in Figure 2d. While this first programmable meta-imager endowed its radiating elements with programmability, it is also possible to endow the interior of the scattering system with programmability. This route was chosen to implement a programmable 3D leaky chaotic cavity[71]: its interior boundaries are partially covered with a programmable metasurface reflect-array[72] but the cavity's radiating slots are static. The reflection properties of each programmable meta-atom can be individually tuned via the bias voltage of a varactor diode. By modifying the scattering inside the cavity using its programmable boundaries, the pseudo-random wave forms that leak out into the far field are altered. More compact leaky 2D chaotic cavities can, of course, also be endowed with programmability of their radiating elements[73] (see Figure 2i) or their interior scattering system[74]. Such programmable leaky chaotic cavities retain, of course, their frequency diversity seen in Figure 2h but their programmability-enabled single-frequency operation frees up frequency resources that can now be deployed to image frequency-dependent scenes; for instance, the programmable 2D leaky chaotic cavity from Figure 2i was used to image a scene containing multiple incoherent sources with both spatial and spectral resolution[75].

In Figure 2f we saw the use of a programmable reflect-array metasurface inside a chaotic cavity to tune the system's scattering properties. Programmable metasurface transmit- or reflect-arrays can also be deployed in free space in conjunction with a carefully aligned horn antenna to generate pseudo-random wave forms for scene illumination[76], as seen in Figure 2k. While this setup may be considered bulky, as pointed out previously, we will see below that it becomes very attractive when combined with stray ambient waves.

*2.2.3 Orthogonal Patterns.* – Up to now, we covered influential examples of compressive meta-imagers that use pseudo-random scene illuminations. The partial redundancy resulting from the finite overlap of subsequent measurements based on a series of pseudo-random patterns motivates the optimization of the measurement modes, and thereby the sensing matrix, toward the use of orthogonal patterns. Numerical studies showed that the diversity of the sensing matrix can be improved by optimizing the position of antennas within a conventional antenna array[77,78], or by optimizing the design of metallic or dielectric "compressive reflector antennas"[79–81], the deployment mode of the latter being similar to previously discussed metasurface reflect-arrays (Figure 2j). Similarly, experiments with programmable metasurface reflect-arrays recently optimized the coding pattern sequence to improve measurement diversity[82,83]. All these works minimize mutual coherence[84–86], that is, they try to flatten the singular value spectrum of the sensing matrix. Usually, they do not achieve perfectly flat singular value spectra, such that some degree of redundancy remains in subsequent measurements. Some of these works make an interesting link between compressive sensing and information theory, by interpreting the sensing matrix as MIMO channel matrix and evoking the generalized Shannon capacity[87–89]. In that perspective, it is important to note that the information capacity depends on both channel diversity and SNR. It is therefore in general not possible to optimize the capacity without fixing a value of SNR. Specifically, a direct correlation between channel diversity and channel capacity in such an optimization procedure can only hold under the assumption that the channel path losses do not change during the optimization. This assumption is tacitly made when normalizing the sensing matrix columns. Incidentally, a conceptually related work on wireless communication in a metasurface-programmable rich-scattering environment implemented perfectly orthogonal channels and observed that the path losses were on average not affected by the optimization procedure[90].

Reports on perfectly orthogonal sensing matrices are rare in the compressive meta-imager literature. An influential work, shown in Figure 2l, demonstrated a programmable metasurface reflect-array in the terahertz regime[91]. The working principle of its constituent pixel relies on a dynamic metamaterial absorber that can be individually tuned with an applied bias voltage. For instance, various patterns from the Hadamard basis can be implement, yielding a series of orthogonal scene illumination patterns without any acquisition of redundant information.

Perfectly orthogonal scene illuminations were also implemented with a programmable 3D leaky chaotic cavity[92], as seen in Figure 2g. The setup resembles that from Figure 2f, except that the programmable boundary of the cavity is not used in random configurations. Instead, the coding sequence is judiciously tailored such that the resulting sensing matrix becomes perfectly orthogonal. Here, the basis is not a well-known basis like the Hadamard basis, and upon visual inspection, the sensing matrix may appear random. However, the overlap between its columns is zero, and its singular value spectrum is perfectly flat. Therefore, using this tailored coding sequences enables one to conduct a series of compressive measurements without any redundancy. Identifying the appropriate coding sequence is non-trivial: due to the complexity of the scattering inside the chaotic cavity, no analytical forward model exists that describes the impact of the coding pattern on the scene illumination. An additional complication is that the optimal configurations of different meta-atoms are correlated due to the reverberation inside the cavity. Ref.[92] iteratively optimized the configuration in a trial-and-error approach.

Before closing this subsection, we note that the arguments for the benefits of orthogonal patterns over pseudo-random patterns in terms of redundancy are valid on average when many completely different scenes are imaged. For one specific scene, or one specific class of scenes, some orthogonal bases perform better than others; the basis that performs best for a given scene class are its principle scene components which we discuss next.

***2.2.4 Principle Scene Components.*** – A special case of orthogonal patterns are principle scene components. Originally pioneered in optical systems[93–95], the idea is to analyze a dataset of representative scenes in order to identify a basis in which a typical scene can be expressed as linear superposition of a small number of "principle" patterns. PCA can be understood as designing the sensing matrix to satisfy the restricted isometry property on the specific dataset of representative scenes[96]. In a scene-agnostic dictionary like the Hadamard basis, some entries may be of very limited use in a given imaging scenario. In contrast, deploying principle scene components expedites the measurement process because the same information can be measured with fewer scene illumination patterns. The trade-off between latency and loss of generality occurs for the first time here and will accompany us throughout the rest of this Review. In many realistic applications the loss of generality does not pose a problem given that they seek to address specific tasks as seen in Figure 1.

A numerical study applied the idea of using principle scene components of human postures as illumination patterns to EM imaging with an array of tunable antennas[43]. An experimental implementation of EM meta-imaging with principle scene components of human postures was recently reported based on a 2-bit programmable reflect-array metasurface[97]. To identify a suitable sequence of coding patterns, Ref.[97] relied on an analytical forward model that describes the programmable metasurface as an array of point-like dipoles whose induced currents depend on the underlying coding pattern, and applied a modified Gerchberg-Saxton algorithm[98]. Moreover, Ref.[97] implemented a reconstruction-free analysis of the measurement data in order to classify the human posture on display in the scene. This idea of skipping the image reconstruction will surface again in the next section.

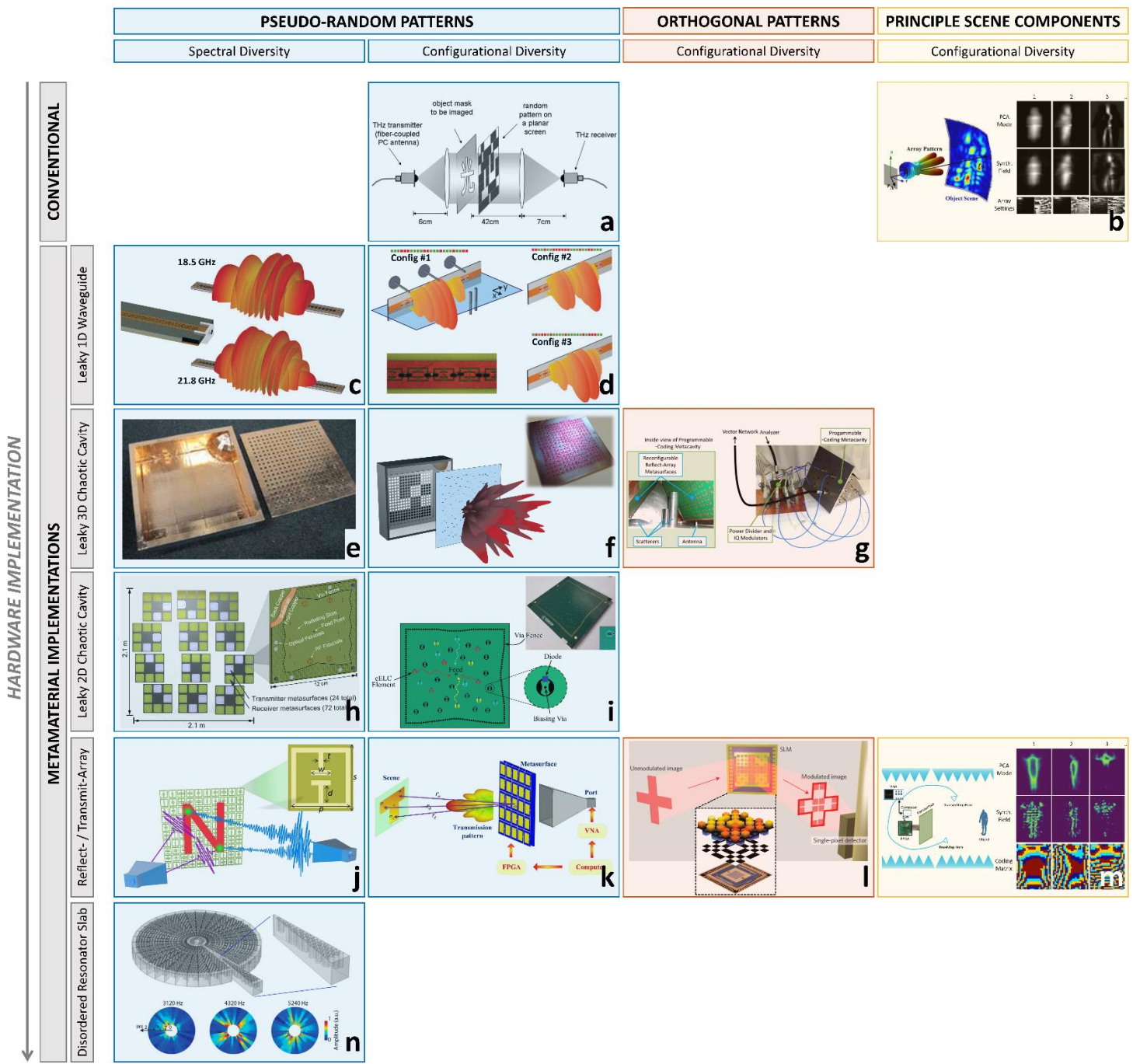

**Figure 2. Overview of compressive meta-imagers in terms of scene illumination strategy and metamaterial hardware implementation.** Compressive scene illuminations can leverage pseudo-random patterns, generated either with spectral diversity (first column) or configurational diversity (second column), orthogonal patterns (third column), or principle scene components (forth column). Orthogonal patterns can be based on a well-known basis (e.g., the Hadamard basis) or any other basis. The hardware implementation of these scene illuminations can follow conventional strategies (first row) or using metamaterial devices. The latter can involve leaky 1D waveguides (second row), leaky 3D chaotic cavities (third row), leaky 2D chaotic cavities (forth row), reflect- or transmit-

arrays (fifth row), or disordered resonator collections (sixth row). (Copyrights: **a,** Reproduced from *Appl. Phys. Lett.* **93**, 121105 (2008),[41] with the permission of AIP Publishing. **b,** Adapted with permission from *IEEE Antennas Wirel. Propag. Lett.* **14**, 1039 (2015).[43] Copyright 2015 Institute of Electrical and Electronics Engineers. **c,** Adapted with permission from *Science* **339**, 310 (2013).[37] Copyright 2013 American Association for the Advancement of Science.of Science. **d,** Adapted with permission from *J. Opt. Soc. Am. B* **33**, 1098 (2016).[99] Copyright 2016 The Optical Society. **e,** Reproduced from *Appl. Phys. Lett.* **106**, 194104 (2015),[46] with the permission of AIP Publishing. **f,** Reproduced with permission from *Phys. Rev. Applied* **6**, 054019 (2016).[71] Copyright 2016 American 2016 American Physical Society. **g,** Reproduced with permission from *Phys. Rev. Applied* **13**, 041004 (2020).[92] Copyright 2016 American 2020 American Physical Society. **h,** Gollub, J. N. *et al.*, *Sci. Rep.* **7**, 42650 (2017);[50] licensed under a Creative Commons Attribution (CC BY) license. **i,** Main: Diebold, A. V. et al., *Optica* **7**, 527 (2020);[75] licensed under a Creative Commons Attribution (CC BY) license. Top right corner: Reproduced with permission from *IEEE Trans. Antennas Propag.* **69**, 2151 (2020).[73] Copyright 2020 Institute of Electrical and Electronics Engineers. **j,** Wang, L. *et al.*, *Sci. Rep.* **6**, 26959 (2016);[54] licensed under a Creative Commons Attribution (CC BY) license). **k,** Li, Y. B. *et al.*, *Sci. Rep.* **6**, 23731 (2016);[76] licensed under a Creative Commons Attribution (CC BY) license). **l,** Adapted with permission from *Nat. Photonics* **8**, 605 (2014).[91] Copyright 2014 Springer Nature. **m,** Li, L. *et al.*, *Nat. Commun.* **10**, 1082 (2019);[97] licensed under a Creative Commons Attribution (CC BY) license). **n,** Reproduced with permission from *Proc. Nat. Acad. Sci. USA* **112**, 10595 (2015).[57] Copyright 2015 National Academy of Sciences.)

**2.3 Image Reconstruction Algorithms.** Before closing the section on compressive meta-imagers, we briefly review popular image reconstruction algorithms. As pointed out in Figure 1, conventional sensing pipelines seek to obtain a visual image of the scene suitable for human consumption before analyzing this image to answer the specific sensing task. Such image reconstruction is inherently more difficult than inference from raw measurements to answer the specific task, implying that unnecessarily many measurements and large computational resources are needed[95]. We will elaborate on the benefits of reconstruction-free inference in the next section on intelligent meta-imagers.

Under the first Born approximation, the inverse scattering problem in computational imaging essentially requires the inversion of the sensing matrix **H** in order to determine an estimate of the scene: $\tilde{S} \approx \mathbf{H}^{-1} D$. Many standard pseudo-inversion techniques such as Tikhonov regularization[100] exist in order to perform this inversion in the presence of noise. Higher reconstruction qualities are often achieved with iterative techniques such as the Generalized Minimal RESidual method (GMRES)[101] or Two-step Iterative Shrinkage/Thresholding (TwIST)[102], at the cost of a higher computation and data storage burden. Recently, Ref.[103] demonstrated that processing time and memory consumption of the previously outlined reconstruction techniques can be improved through truncation of insignificant principal components of the sensing matrix via a PCA analysis of the latter; this technique is not to be confused with the use of PCA to identify principle scene components as illumination patterns in the previous subsection 2.2.4[43,97]. Another attempt at faster processing showed that compressive meta-imaging is compatible with fast range migration algorithms originating from synthetic-aperture-radar (SAR) imaging[104–106] that take advantage of efficient fast Fourier transforms[107–109]. Compressive meta-imaging can also be applied to SAR, that is imaging with a moving meta-imager. To that end, the functionality of well-known SAR modalities (spotlight and stripmap illumination) can be achieved through pattern synthesis with programmable meta-imagers, but stripmap illumination was also implemented with pseudo-random patterns[110].

In terms of hardware cost, it is often advantageous to rely on intensity-only measurements, especially at higher frequencies or in specific applications like bistatic space-born SAR[111]: phaseless acquisition removes the need for phase synchronization between transmitter and receiver[112], minimizes the requirements for phase stability of the local oscillator, and limits effects of calibration errors. One solution to deal with phaseless data is to apply phase-retrieval techniques[113,114], as was done in the context of compressive meta-imagers in Refs.[115–117]. An alternative to phase retrieval techniques is found in computational "ghost imaging"[118–120] which correlates the intensity measurements with the intensity of the scene illumination patterns. This technique has been successfully applied to compressive meta-imagers in Refs.[121–123] and also served as the basis of an optical encryption scheme in combination with helicity-dependent meta-holograms[124]. A phaseless computational imaging method that has not yet been applied to compressive meta-imaging is ptychography[125] which generates a high-resolution image from a series of low-resolution intensity measurements with structured scene illumination.

More recently, various machine learning and deep learning algorithms have been applied in computational imaging[126–130] and inverse scattering[131,132], including some cases of compressive meta-imagers[97,133,58,82,134,83,135]. Their potential benefits include fast online inference on tasks that are difficult or impossible to formulate analytically, at the cost of expensive offline training accompanied by a need for a large training dataset. With reference to the sensing pipeline from Figure 1, such AI tools have been used in compressive meta-imagers at the image reconstruction stage in Ref.[133,58,83] and to interpret the resulting image in order to answer the specific task in Refs.[97,58,135,83]. Moreover, AI tools have been used in Refs.[82,97,134] to interpret raw data obtained with compressive meta-imagers without reconstructing an image; we will return to this concept of reconstruction-free inference in more detail in section 3.2. However, as pointed out previously, none of these cases fully reaps the benefits of novel AI tools in order to optimize the *physical* measurement process, which is why we do not refer to these works as "intelligent" meta-imagers in this Review.

## 3. Intelligent Meta-Imagers

All of the compressive meta-imagers covered in the previous section illuminate the scene without taking into account the ultimate purpose of the sensing procedure. In other words, the measurements are agnostic to the task. In this section, we turn our attention to "intelligent" meta-imagers which sculpt the scene illumination in accordance with the specific sensing task. The role of AI in these systems is hence not restricted to *digital* data processing but actively influences the *physical* measurement process. We refer to a meta-imager as "intelligent" if AI controls the configuration of its programmable meta-hardware such that the measurement process becomes task-aware and thus purposefully non-isometric. We thereby witness how the physical world of metamaterial engineering and the digital world of AI merge in the context of imaging and sensing. In related fields, "self-adaptive" intelligent metasurfaces have recently been studied for dynamic beam steering[136] and dynamic cloaking[137].

To begin, we cover the autonomous identification and analysis of a region-of-interest (ROI) within the scene. Next, we analyze the flow of information through the sensing pipeline and highlight the fundamental difference between compressive and intelligent meta-imagers: the latter discriminate between task-relevant and task-irrelevant information already during the measurement process. In other words, while compressive meta-imagers seek an isometric projection to faithfully compress all scene information, intelligent meta-imagers process the scene information on the physical layer with respect to the specific task by purposefully distorting the distances between points in a non-isometric projection. Then, we detail how the underlying "learned sensing" strategy can be implemented in practice by covering two recent implementations from the literature.

**3.1 Autonomous Region-of-Interest Identification and Analysis.** In advanced EM imaging and sensing tasks it is oftentimes helpful to identify the region of interest (ROI) within a very large scene (compared to the wavelength). On the one hand, the analysis can therefore be restricted to the ROI as opposed to the entire scene. In Ref.[50], the meta-imager was coupled with an optical light sensor (Kinect); this sensor fusion enabled an automatic identification of the ROI which was used to limit the microwave-image reconstruction burden. However, the ROI identification did not influence the frequency-diverse pseudo-random scene illumination patterns such that we regard this example as a compressive rather than intelligent meta-imager.

On the other hand, ROI identification can also be important to enable subsequent illumination of this area of particular interest in the scene for analysis with improved resolution or SNR. In the context of optical ghost imaging, in Refs.[138–140], a simple algorithm identifies a ROI (based on its high spatial frequency content[138,139] or motion[140]) and subsequently chooses illumination patterns to analyze the ROI with higher resolution. In these examples, ROI identification and illumination pattern synthesis are performed analytically without resorting to AI tools and the ultimate goal is image reconstruction rather than answering a specific sensing task. In the context of microwave meta-imaging, in Ref.[133], first, a convolutional neural network performs image reconstruction based on measurements obtained with pseudo-random configurational diversity; second, based on the reconstructed image from the first step, another neural network identifies the hand and chest areas of the imaged person; third, given the inferred spatial position of these ROIs, the programmable meta-imager synthesizes illumination patterns to focus on the ROIs; forth, a third neural network interprets the measured data from this second series of measurements, for instance, to infer details about the displayed hand gesture. In this complex sequence of sensing tasks, illustrated in Figure 3, AI determined the configuration of the programmable meta-imager in the second series of measurements; by focusing on the ROIs, significant SNR enhancements for tackling the specific task were achieved. This work thus constitutes a first example of an intelligent meta-imager, even though its performance could be further improved by using intelligent as opposed to pseudo-random illumination patterns during the first measurement step. We discuss how intelligent patterns can be implement in subsection 3.3.

Another important feature in Ref.[133] is related to its hardware: the programmable meta-imager is based on a programmable metasurface reflect-array. Other works with such hardware (see Figure 2j,k,m)[54,76,82,83,97] relied on illumination with a carefully aligned horn antenna, resulting in an overall bulky setup. In contrast, Ref.[133] demonstrates the above-outlined functionalities also if stray ambient WiFi waves serve as (uncontrolled) source of illumination. This result paves the way toward mounting programmable metasurfaces on walls and opportunistically using ambient radiowaves for advanced sensing tasks. Not only does this approach resolve the bulkiness concern, but it also recycles waves instead of emitting additional ones.

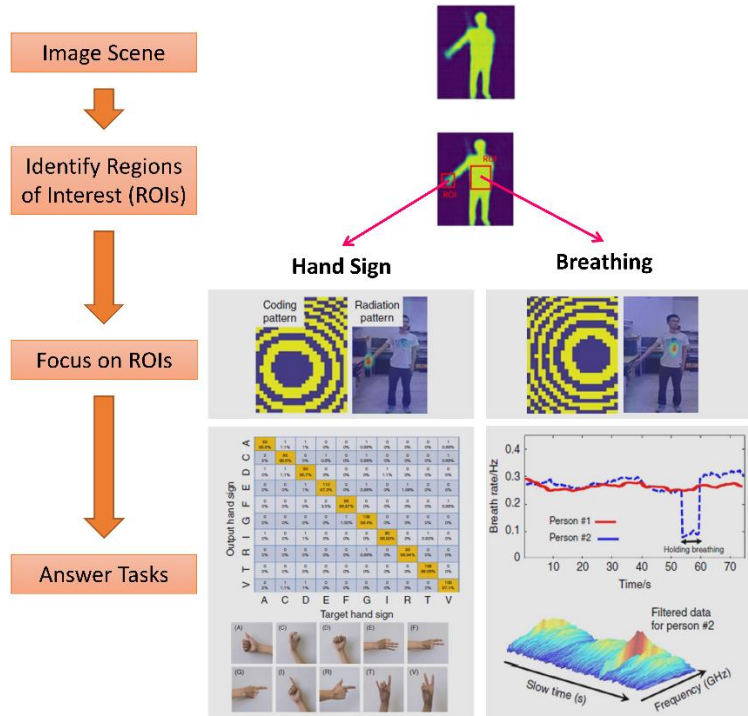

**Figure 3. Machine-learning-controlled meta-imager autonomously identifies regions of interest and performs multiple complex tasks.** First, the scene is imaged using pseudo-random patterns and a neural network for image reconstruction. Second, the image is analyzed to identify regions of interest for analyzing hand signs and breathing. Third, the metasurface configuration is adapted to focus sequentially on the identified regions of interest. Forth, the acquired data is processed, in some cases using a neural network, to answer the tasks at hand. The panels are adapted with permission from Li, L. *et al.*, *Light Sci. Appl.* **8**, 97 (2019);[133] licensed under a Creative Commons Attribution (CC BY) license.

**3.2 Information Flow and Operation Principle.** To prepare the ground for the concept of intelligent computational meta-imaging, we start by retaking the sensing pipeline from Figure 1 in Figure 4a in order to analyze the flow of information through the sensing pipeline in terms of its relevance to the specific sensing task. The conventional sensing pipeline deployed in compressive meta-imagers (Figure 4a) drags task-irrelevant information from the scene into the data (because the measurements are task-agnostic) and further into the reconstructed image (because the image reconstruction is also task-agnostic). The task-relevant information is separated from the task-irrelevant information only at the very last stage of image interpretation. Dragging irrelevant information through the sensing pipeline results in numerous inconveniences related to the need for more measurements and a larger dataset to be processed. As a result, latency is unnecessarily poor, as are other metrics like power consumption, data storage, radiation exposure and computational-power requirements. An intelligent sensing pipeline would discriminate between relevant and irrelevant information at the earliest possible stage (the measurement) and skip the image reconstruction step, as sketched in Figure 4b. Thereby, the measurement process of intelligent meta-imagers is simultaneously an analog wave processor that implements a first task-specific inference step. Consequently, no clear distinction between measurement and interpretation is possible in the case of intelligent meta-imagers.

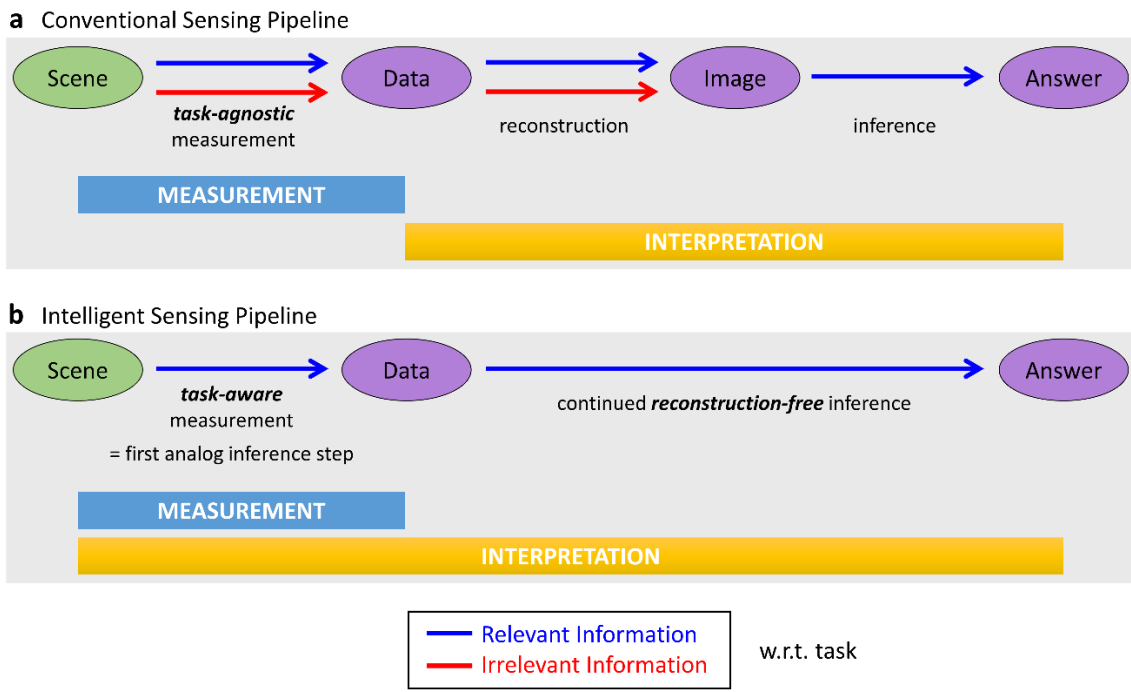

**Figure 4. Flow of (ir)relevant information through the sensing pipeline. a,** A conventional sensing pipeline is based on a task-agnostic measurement process and drags irrelevant information (red) from the scene through the data to the image, only discriminating between relevant and irrelevant at the very last stage. **b,** In contrast, an intelligent sensing pipeline selects only relevant information already

during the physical measurement process. In an intelligent sensing pipeline, the measurement process is therefore simultaneously a first analog inference step. Moreover, it skips image reconstruction and directly interprets the data to answer the task at hand.

A first improvement, purely on the digital layer, is thus an idea that we already encountered in a few examples: reconstruction-free interpretation of raw data[95,141,142]. The latter logically follows from the realization that in the vast majority of sensing scenarios, the ultimate goal is to answer a specific task rather than to obtain a visual image of the scene suitable for human consumption. Most such inference tasks are inherently simpler than image reconstruction, fundamentally, because not all scene information is relevant to them. In terms of the flow of (ir)relevant information through the sensing pipeline, the discrimination between relevant and irrelevant thus happens directly *after* the measurement process. More formally, the reason why inference is possible directly in the compressed latent domain is that, according to the Johnson-Lindenstrauss lemma, the degree of correlation between two signals is approximately the same after compression[95,143]. Some recent works on compressive meta-imagers follow the idea of reconstruction-free inference and deploy neural networks to directly interpret the raw measurement data with respect to the specific sensing task[82,97,134]. However, purely reconstruction-free sensing still acquires irrelevant information in a task-agnostic measurement process; hence the physical layer in these cases lacks intelligent task-awareness and we regard them as examples of compressive rather than intelligent meta-imagers.

A generic intelligent sensing pipeline, as seen in Figure 4b, goes significantly beyond reconstruction-free inference: a task-aware measurement process discriminates between relevant and irrelevant scene information already during the measurement. An intelligent measurement is *not* a faithful analog compressor of *all* scene information. Instead, it promotes features relevant to the task at hand and simultaneously suppresses irrelevant features. The sensing matrix should thus act as a *physical* discriminator between relevant and irrelevant information.

A related idea has been well understood in the signal processing community for more than two decades. A common problem there is robust face recognition based on a digital photographic image despite variations in lighting and facial expression. To that end, the image is linearly projected into a lower-dimensional subspace. One possible projection is based on PCA: such a PCA-based projection reduces the dimensionality almost without distorting the relative distance between points, implying that unwanted variations due to lighting and facial expressions are retained. PCA projections maximize the total scatter across the entire dataset because they are unaware of which class different examples belong to. Therefore, while being optimal for data compression, PCA patterns are not optimal for discrimination between classes as needed for inference. Alternatively, the projections can be based on a Fisher's linear discriminant analysis (LDA)[144] which maximizes the distance between points belonging to different classes while minimizing the distance between points belonging to the same classes[145]. In other words, unwanted variations are discounted in such an LDA-based projection. Of course, the success of LDA over PCA requires a sufficiently large and representative training dataset[146]. In signal processing theory, both PCA and LDA are covered by the broad term "dictionary learning"[147,148], a terminology that blurs the distinction between near-isometric and purposefully non-isometric linear embeddings. For this reason, we avoid using the term "dictionary learning" in the present paper. While the described signal processing problem of face recognition introduces the interest of purposefully non-isometric task-specific linear embeddings[145,149,150], it differs from our application in EM sensing in two ways. First, and most importantly, the task-aware projection happens on the digital layer as part of the data analysis whereas in our EM sensing applications it should be implemented on the physical layer as part of the measurement process via intelligent scene illumination patterns. Second, the outcome of the projection, that is the measured data in our case, is further processed in our EM sensing pipeline by a neural network involving non-linear operations.

More recently, such task-specific projections have been implemented physically in optics using, for instance, digital micromirror devices, to illuminate the scene[95,151]. With the proliferation of AI and deep learning tools, it has become clear that if a neural network is used to analyze the measured data, one may as well include the programmable measurement hardware as additional first layer in the neural network. This results in a hybrid analog-digital end-to-end sensing pipeline containing both trainable physical weights and trainable digital weights that can be jointly optimized. Thereby, the optimization of the physical weights takes into account the specific task, irrespective of how complex and non-linear the digital processing part is. Early works applied this idea to optimizing the color multiplexing pattern of a camera to reconstruct a full color image[152] and the illumination arrangement of a microscope to detect malaria-infected cells[153]. Similar approaches were subsequently studied in many works in the optical domain[154–166], in the ultrasonic domain[167], as well as for meta-imaging in the EM domain[168,169]. The same idea of learned sensing has recently also been proposed generically as data-driven computational sensor design without any concrete implementation in Ref.[170].

In general, in all these works there is no guarantee that the identified optimized physical layer is globally optimal, as in most optimization problems. However, experience with EM intelligent meta-imagers[168,169] typically shows that different optimization runs yield different local optima of roughly similar quality (see, for instance, Figure 6a below). Some ideas to get closer to the global optimum have been explored in the context of nanophotonic inverse design[171]. Provably optimal coherent illumination patterns have, to date, only been identified for (i) binary decision problems[172], or (ii) precisely estimating small (perturbative) variations in the value of one continuous parameter[173]. In the former case, it is clear from linear algebra that the optimal wavefront to discriminate between two system configurations is the first eigenstate of $(\Delta T)^\dagger (\Delta T)$, where $\Delta T$ is the difference between the transmission matrices for the two configurations to be distinguished and $\dagger$ denotes the conjugate transpose. In the latter case, the same approach is applied in the limit of small perturbations of one parameter $\theta$ around some fixed value $\bar{\theta}$, such that illumination with the first eigenstate of $(\partial_\theta T)^\dagger (\partial_\theta T)$ maximizes the impact of changes in $\theta$ on the output field, as long as $\theta$ remains in the vicinity of $\bar{\theta}$. Interestingly, these operator-
29 / 63

based approaches are valid irrespective of whether the scene is in free space or embedded in a complex scattering environment; however, whether it is possible to scale up such operator-based approaches to non-perturbative variations of many parameters remains an open question.

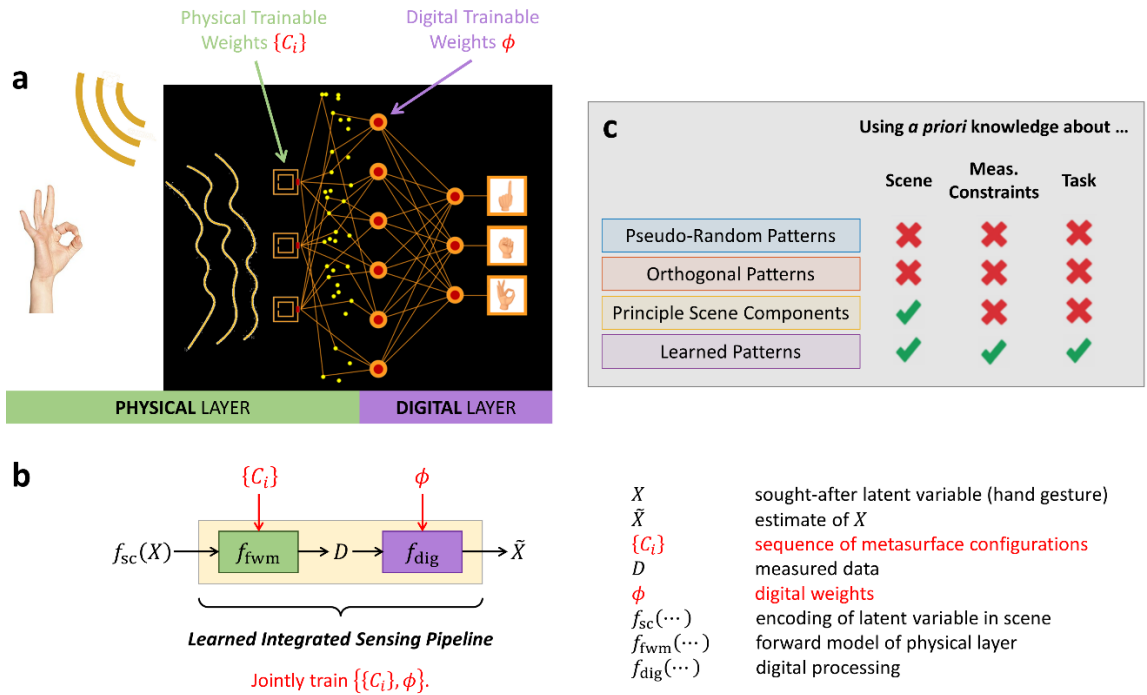

**Figure 5. Operational principle of learned-sensing intelligent meta-imagers. a,** Concept: an end-to-end sensing pipeline contains trainable physical weights (programmable meta-atoms) and trainable digital weights (neural network nodes). The figure is adapted with permission from del Hougne, P. From compressed sensing to learned sensing with metasurface imagers. in *Proc. Passive and Active Millimeter-Wave Imaging XXIV* (eds. Robertson, D. A. & Wikner, D. A.) (SPIE, 2021).[174] Copyright 2021 SPIE – the international society for optics and photonics. **b,** The learned integrated sensing pipeline maps the scene, a function of the sought-after latent variable $X$, to an estimate of $X$ via a physical and a subsequent digital layer. The configurations $\{C_i\}$ of the programmable meta-atoms during the measurement process and the weights $\phi$ of the digital layer are jointly trained. **c,** Taxonomy of scene illumination strategies in terms of the *a priori* knowledge taken into account: only the learned patterns are task-aware. The figure is reproduced with permission from del Hougne, P. *et al.*, *Adv. Sci.* **7**, 1901913 (2019);[168] licensed under a Creative Commons Attribution (CC BY) license.

Figure 5a summarizes the general concept of intelligent meta-imaging for a setup with metamaterial receiving hardware. The scene scatters a wave towards the metamaterial-based detector. The detection process involves scattering of the wave both by programmable meta-atoms (illustrated as diode-tunable split-ring resonators in the figure) and other fixed scatterers. The information flow then seamlessly continues on the digital layer, where the information scatters through programmable digital nodes until the final layer outputs an answer to the specific task. The hybrid end-to-end pipeline thus contains both physical and digital programmable weights that can be trained jointly.

In terms of information flow, the entire pipeline can be understood as an approximation of the identify function, as seen in Figure 5b. The sought-after information $X$, for instance, the class of a hand gesture, is encoded in the scene reflectivity $S = f_{sc}(X)$ via some function $f_{sc}$. The learned integrated sensing pipeline (LISP) consists of the physical layer parametrized by the coding sequence $\{C_i\}$ of the programmable meta-atoms and the digital layer parametrized by digital trainable weights $\phi$. The LISP's output is an estimate $\tilde{X}$ of the sought-after information $X$. This mapping of $X$ to itself is reminiscent of an auto-encoder (see Ch. 14 in Ref.[175]) except that it lacks the common architectural symmetry of auto-encoders and, most importantly, that the encoding step is implemented physically through wave propagation – as opposed to digitally in the well-known all-digital auto-encoders[176]. Moreover, the objective is a different one: we seek to jointly optimize measurement and processing to answer a specific task whereas conventional auto-encoders are used in data analysis for dimensionality reduction and/or denoising[177].

The use of a single end-to-end pipeline implies that the physical weights are programmed in a way that considers not only *a priori* knowledge about the scene as in principle scene components (subsection 2.2.4) but additionally *a priori* knowledge about the task and potential measurement constraints. This is yet another way to understand why learned patterns outperform compressive patterns (pseudo-random, orthogonal, principal scene components), in line with the previous argument in terms of eliminating irrelevant information already during the measurement process through purposefully non-isometric projections. Wave propagation on the physical layer thus performs task-specific analog signal processing in the case of intelligent meta-imagers: relevant information is highlighted and compressed. In contrast, compressive meta-imagers merely compress all information without performing any selection of salient features.

The use of learned illumination patterns does *not* imply any additional overhead during runtime because, similarly to all compressive illumination techniques, a fixed series of patterns is used that is determined before runtime. Similarly to the use of orthogonal patterns and principle scene components, the use of learned patterns requires a one-off preparatory step before runtime to determine these patterns. Learning an intelligent fixed sequence of illumination patterns requires the availability of sufficient labelled training data; note that identifying a series of principle scene components also requires training data albeit there is no need for labels. The need for labelled training data arises whenever AI is used, be it only for data interpretation based on compressive measurements as in Refs.[97,133,58,82,134,83,135], or for jointly learning intelligent illumination patterns and a matching digital layer as in Refs.[168,169].

**3.3 Implementations.** To illustrate how the concept of intelligent meta-imaging can be implemented in practice, we now closely look at the two existing examples from the literature[168,169]. The first demonstration of an intelligent learned-sensing meta-imager was a numerical study in which transmitter and receiver each take the form of a dynamic metasurface transceiver (similar to Figure 2i), aiming to recognize a metallic digit in free space – see Figure 6a[168]. Subsequently, the first experimental implementation of learned sensing with meta-imagers was reported – see Figure 6b – where the scene is illuminated with waves generated by a horn antenna and reflected off a programmable metasurface reflect-array, aiming to recognize a human posture[169].

Both implementations of learned-sensing intelligent meta-imagers achieved remarkable latency improvements compared to the compressive sensing approaches from section 2. In particular, Ref.[168] benchmarked the average achievable classification accuracy as a function of the number of utilized scene illumination patterns against compressive meta-imaging paradigms. Obviously, the lower the number of utilized scene illumination patterns, the better is the system latency. Orthogonal patterns are seen on the right in Figure 6a to yield a very small advantage over pseudo-random patterns as the number of utilized scene illuminations increases, because redundancy in subsequent measurements is eliminated. Principal scene components yield a more notable improvement over orthogonal patterns, because they are optimal to compress the expected scene information. However, note that the synthesis of orthogonal patterns and principle scene components is in general not perfectly implemented with constrained hardware. Learned patterns yield a remarkably clear accuracy improvement over all three compressive sensing benchmarks, especially for low-latency systems in which the number of scene illuminations is very limited. This superior performance can be attributed to the fact that learned patterns take into account the specific task and measurement hardware constraints (see also Figure 5c). Note that using non-random patterns (orthogonal patterns, principle scene components, learned patterns) does not imply an additional computational or latency cost at runtime because these non-random patterns are determined offline in a preparatory step before runtime.

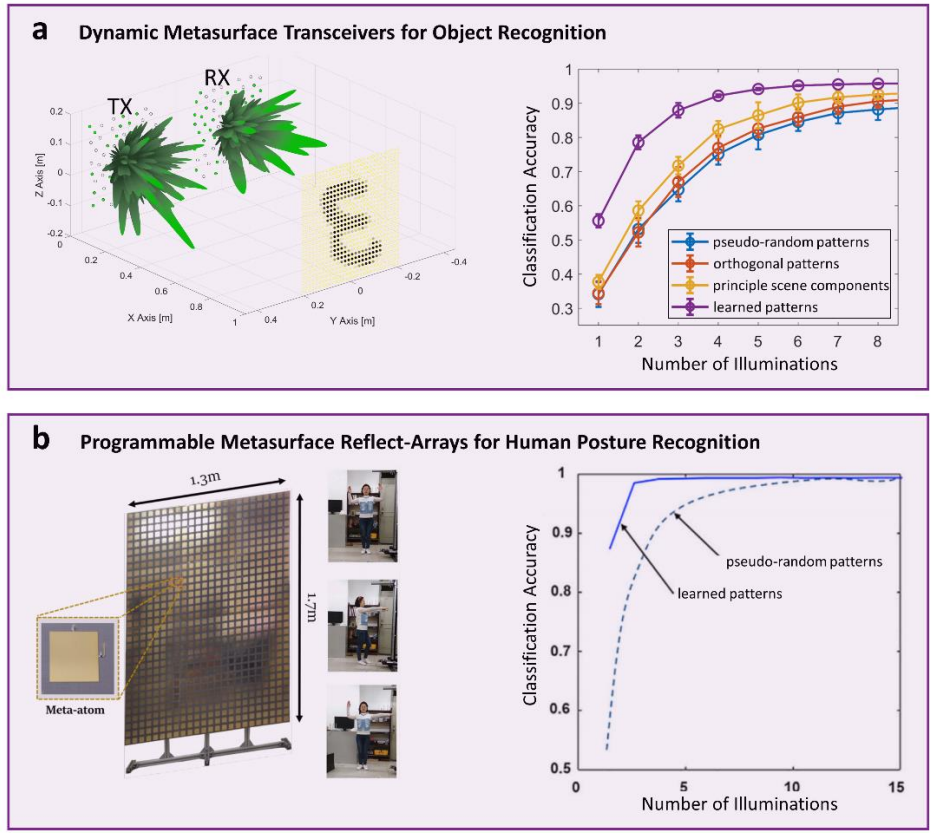

**Figure 6. Implementations of intelligent meta-imagers. a,** Numerical demonstration of an intelligent meta-imager used to recognize metallic objects. The meta-device is based on a 2D leaky waveguide. Its classification accuracy clearly outperforms the three state-of-the-art compressed sensing benchmarks (see Figure 2), especially when the number of patterns is limited (low latency). The panels are adapted with permission from del Hougne, P. *et al.*, *Adv. Sci.* **7**, 1901913 (2019);[168] licensed under a Creative Commons Attribution (CC BY) license. **b,** Experimental demonstration of an intelligent meta-imager used to recognize human gestures. The meta-device is based on a programmable reflect-array metasurface illuminated by a horn antenna. Similar latency improvements are observed. The panels are adapted with permission from Li, H.-Y. *et al.*, *Patterns* **1**, 100006 (2020);[169] licensed under a Creative Commons Attribution (CC BY) license.

In the following, we break down the five major challenges in realizing a learned-sensing pipeline with programmable metamaterial hardware, and we detail the distinct ways in which Refs.[168,169] have mastered them.

*3.3.1 Forward Model.* – The first step toward assembling a learned-sensing pipeline, as seen in Figure 5b, is to establish a forward model $f_{fwm}(f_{sc}(X), \{C_i\})$ of the physical layer that maps a given scene $f_{sc}(X)$ to the measured data $D$ as a function of the chosen sequence of coding patterns $\{C_i\}$ for the programmable meta-atoms. This forward model must be differentiable so that, once combined with the subsequent digital layers, the hybrid neural network can be trained through error back-propagation. In some cases, it is feasible to conceive an analytical forward model. For instance, in Ref.[168] an efficient compact description of the radiating sub-wavelength meta-atoms as dipoles is used that takes into account coupling effects[178]. Analytical models based on such discrete-dipole approximations have successfully been verified for stacked 1D leaky waveguides with programmable radiators in experiments[179], as well as for 2D metasurface transceivers in full-wave simulations and experiments[180,181]. Simple analytical models of programmable metasurface reflect-arrays also yielded good experimental results, for instance, in Refs.[97,98]. In other cases, however, an analytical description of the forward model may be too cumbersome, non-differentiable or unfeasible to obtain. An example of the latter would be hardware such as the 3D leaky programmable chaotic cavities[71] from Figure 2f. In such cases, a neural network can be trained to approximate the physical forward function $f_{fwm}$, as was done in Ref.[169]. In that case, an auxiliary neural network that maps $f_{sc}(X)$ and $\{C_i\}$ to $D$ is trained and, once trained, its internal weights are kept fixed. A learned forward model is guaranteed to be differentiable. Once a forward model, be it of analytical or learned nature, has been obtained, it can be integrated into the hybrid sensing pipeline from Figure 5b.

*3.3.2 Digital Neural Network Architecture.* – The use of convolutional architectures is very popular[153,133,169] because in the field of computer vision they displayed an excellent ability to identify local correlations in data. However, unlike computer vision problems, we are dealing with multiplexed data such that one may hypothesize that local features are actually encoded in long-range correlations. Therefore, surprisingly simple fully connected layers are often a good choice for the digital neural network[69,168,182–184].

*3.3.3 Training Data.* – The next step is to accumulate enough data to train the end-to-end hybrid neural network seen in Figure 5b, containing both physical and digital layer. For supervised learning, usually a reasonably large number of labelled examples is required. In our case, this corresponds to pairs of a scene $f_{sc}(X)$ and its corresponding label $X$. For instance, if we are interested in classifying hand gestures, $f_{sc}(X)$ would be a map of scene reflectivity and $X$ would identify the corresponding hand gesture. The approach taken by Refs.[133,169] relied on a commercial digital optical camera, integrated with the programmable metasurface, that can rapidly capture a large amount of labelled optical images of, for instance, hand gestures. After background removal and binarization, these visual images were assumed to approximately correspond to the scene reflectivity in the microwave regime. Another approach could be to rely on synthesized data from simulations that faithfully resemble the experiment. The hybrid neural network could be initially trained on the synthesized data and subsequently be fine-tuned with data from a few experimental microwave measurements. This approach is known as transfer learning. Finally, we point out that various data augmentation techniques also exist in the deep learning literature that may help to diversify small datasets.

*3.3.4 Back-Propagation Under Quantization Constraints.* – Finally, it is time to *jointly* train the physical and digital weights. Many efficient algorithms for error back-propagation[185–187,175], such as Adam[188], are openly accessible. However, usually the programmability of the physical weights is quantized rather than continuous, in most cases restricted to one bit. This quantization appears at first sight incompatible with the need for continuous tunability to implement error back-propagation. Fortunately, there are various tricks to combine the quantization constraints of programmable meta-atoms with error back-propagation:

(i) The so-called "temperature parameter" trick used in Ref.[168] was introduced in Ref.[152] in order to gradually drive the distribution of a given physical parameter $n$ from continuous to discrete. Let $N = [n_1, n_2, \ldots, n_p]$ be a vector collection all $p$ possible discrete values of $n$, $W = [w_1, w_2, ..., w_p]$ an auxiliary weight vector of equal size, and $\beta$ a scale factor that increases with the number of iterations. The variable of interest $n$ can then be expressed as $n = \sum_{j=1}^{p} w'_j n_j$, where $W' = f_{\text{SoftMax}}(\beta|W|)$. As $\beta \to \infty$, $n$ will take one of the allowed discrete values.

(ii) The so-called randomized simultaneous perturbation stochastic approximation (r-SPSA) algorithm used in Ref.[169] originates from petroleum engineering[189] and tests at every iteration whether flipping the state of a randomly selected subset of meta-atoms improves the optimization metric.

(iii) A further possibility is the use of projected gradient descent[190] which treats the physical weights as continuous but projects them onto the allowed discrete values after every $k$th iteration.

*3.3.5 Noise.* – The literature on meta-imaging usually tacitly assumes operation in the high-SNR regime where the impact of detector noise is negligible. Whereas the compressive patterns from section 2 are determined irrespective of measurement constraints, learned patterns are expected to vary with SNR. Specifically, in the very low SNR regime, one may expect different learned patterns that, for instance, highlight the subset of salient features that can be measured with a higher signal strength than others. Applications of intelligent meta-imagers in the low-SNR regime have to date not been explored but may be relevant to space applications with low photon counts[191] as well as to indoor applications recycling weak stray ambient WiFi waves (e.g., Ref.[133]). Incidentally, Ref.[133] performed focusing in the second step to improve the signal strength reflected from the ROI, however, the choice to focus waves on the ROI, as opposed to using some other possibly non-intuitive pattern, was made by the human operator rather than AI.

## 4. Future Directions

At this stage, the reader may wonder: what is next? Do the learned-sensing intelligent meta-imagers from the previous section fully reap the benefits of AI and programmable metamaterials? Our answer is: no, there is room for improvement! The previously reviewed evolution from compressive to intelligent meta-imagers has already enabled remarkable latency improvements but several milestones toward the holy grail of an ultrahigh-resolution cognitive all-analog meta-imager remain. First, we anticipate that future meta-imagers will be endowed with a cognitive feature, allowing them to adapt their programmable metamaterial layer on-the-fly to the information that they measure. Second, we anticipate significant resolution improvements, potentially well below the diffraction limit, by realizing and reaping the benefits of operation in closed reverberant environments as opposed to free space, which is a natural setting for many applications anyway. Third, we anticipate that the entire digital part of the hybrid sensing pipeline from Figure 5a will be implemented with wave-based analog signal processing in the native (microwave) signal domain, yielding substantial improvements in speed and energy consumption. We detail these three future directions in the following.

**4.1 Cognitive Meta-Imagers.** One way to understand the remarkable latency improvements of intelligent meta-imagers over compressive meta-imagers is related to the fact that they make use of seemingly all available *a priori* knowledge on scene, measurement constraints, and task, as seen in Figure 5c. Such intelligent meta-imagers operate with a *fixed* learned sequence of coding patterns (and hence scene illumination patterns). In principle, a *flexible* sequence of coding patterns in which each pattern is adapted based on the information measured with previous patterns in the sequence offers the possibility to make use of an additional source of *a priori* knowledge: the information contained in the data measured with the previous patterns in the current coding sequency. For instance, if the goal is to distinguish between five hand gestures, after a few measurements it may already be clear that with overwhelming probability the hand gesture must be one of two, and it would be more efficient to then use scene illuminations that discriminate specifically between the remaining two candidates. In fact, the number of measurements itself would become flexible and the cognitive meta-imager would decide itself at what point it has collected enough information to answer the specific task with high accuracy.

Incidentally, this approach is chosen by nature, for instance, in the way in which humans selectively focus their visual attention on different parts of the visual space in subsequent glimpses, and has inspired reinforcement learning methods in computer vision[192]. Precursors of cognitive (self-adaptive) computational imaging are ideas related to autonomous ROI identification and analysis from section 3.1. For example, in the area of optical ghost imaging, acquisition algorithms were proposed that sequentially focus their attention on areas with high spatial frequency content[138,139] or motion[140] and that decide themselves when enough measurements have been taken; however, these examples were still aimed at image reconstruction as opposed to answering a specific sensing task for which the optimal illumination patterns cannot be determined intuitively. A first attempt at implementing task-specific cognitive imaging was recently reported in optical microscopy where a biological sample is illuminated with a programmable LED array in order to detect malaria[193] – see Figure 7. Transposing this cognitive imaging paradigm to programmable metamaterial hardware will herald the era of cognitive metamaterials which are capable of self-learning how they should be configured[194].

The conditions under which the *cognitive* meta-imaging framework discussed in this subsection may outperform the *learned* meta-imaging paradigm (Section 3) remain to be determined in future work. Cognitive meta-imaging involves feedback *during* runtime which implies additional overhead that could reduce or eliminate potential advantages. In contrast, the learned meta-imaging paradigm from Section 3 does *not* involve any additional overhead during runtime because it uses a fixed sequence of illumination patterns that were learned for the specific task *beforehand* in a one-off preparatory step.

39 / 63

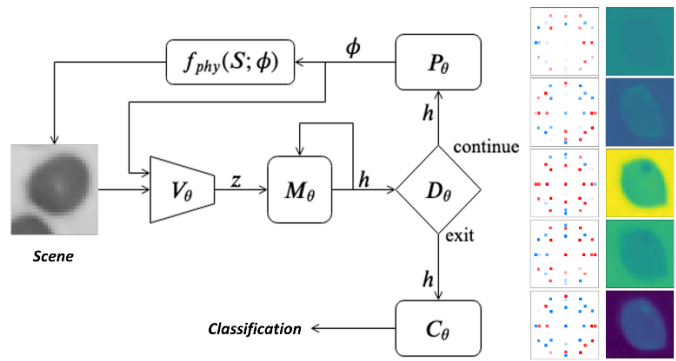

**Figure 7. Cognitive imaging.** Instead of using a fixed series of learned patterns, the imager adapts each pattern based on the information measured with previous patterns in the sequence. The concept is illustrated for an optical microscope in which an LED array self-adaptively illuminates a biological sample to detect malaria. The figure is adapted with permission from Chaware, A. *et al.* Towards an Intelligent Microscope: Adaptively Learned Illumination for Optimal Sample Classification. in *Proc. ICASSP* 9284 (IEEE, 2020).[193] Copyright 2020 Institute of Electrical and Electronics Engineers.

**4.2 Meta-Programmable Reverberation-Coded Apertures.** Besides latency, another roadblock for current EM meta-imager applications is resolution: the wavelength of WiFi waves, for instance, is comparable to the size of a human hand whose gesture may have to be determined based on recycled WiFi waves (as in Ref.[133]). We have previously pointed out that range resolution is limited by the utilized bandwidth *only if* no waves are incident at oblique angles[19], resolving the apparent paradox of achieving range resolution with single-frequency measurements in Refs.[67,68,195]. Pushing this idea further, one may wonder what resolution is achievable when the entire scene is enclosed by reflecting walls, resulting in secondary waves impinging on the scene from all possible angles. In this subsection, we discuss this special type of coded aperture, the reverberation-coded aperture (RCA)[69], and explain why it is capable of compressing deeply sub-wavelength scene information without access to evanescent waves. RCAs can leverage spectral or configurational diversity, similar to conventional coded apertures in Figure 2. Upgrading intelligent meta-imagers to RCA hardware may be as simple as moving them from contrived anechoic test chambers to real-life indoor environments. In fact, many practical scenarios already operate inside enclosed reverberant environments, usually without realizing the implications thereof regarding the achievable resolution. Examples include indoor localization[196–199,183,200] and imaging[201], biomedical breast imaging inside (reconfigurable) metallic enclosures[202], impact localization with elastic waves propagating on solid surfaces[203,204], or photo-acoustic imaging inside an acoustically reverberant cavity[205]. Many applications of EM meta-imaging indeed involve operation in indoor environments which naturally act as RCAs, many of which will soon be endowed with programmability as part of next-generation wireless communication paradigms[206,207,200].

While the scene is *outside* of a conventional coded aperture that relies on the spectral or configurational diversity of a leaky chaotic cavity, in an RCA the scene is *inside* the cavity – see Figure 8a-c. This apparently small difference has large consequences: the wave interacts with the scene not once but countless times such that it develops a much stronger sensitivity to sub-wavelength scene details[69]. It is instructive to think of the RCA as a "generalized interferometer". Indeed, it is well known that the interferometric sensitivity, for instance for a high-finesse Fabry-Perot cavity[208], is typically deeply sub-wavelength – without any access to evanescent waves nor any link to wave focusing. While basic optical interferometers are essentially one-dimensional devices, RCAs generalize this sensitivity to two or three dimensions, making the interferometric sensitivity exploitable for complicated sensing tasks. Of course, the first Born approximation is not satisfied in an RCA but since we seek reconstruction-free inference, the raw multiplexed data can be directly interpreted by a neural network.

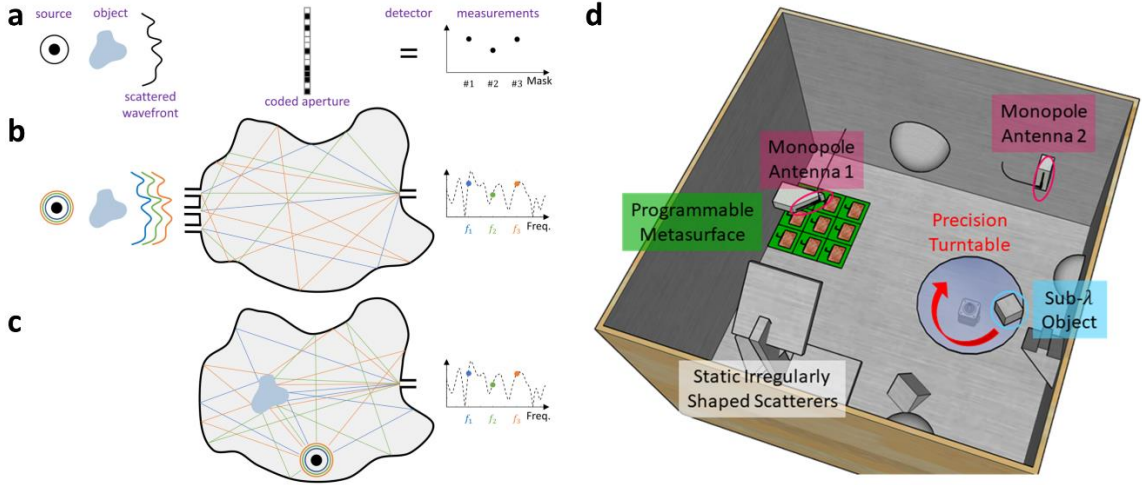

**Figure 8. Meta-Programmable Reverberation-Coded Aperture.** Significant resolution improvements can be achieved by placing the scene inside a complex enclosure such that reverberating waves develop a much higher sensitivity to sub-wavelength scene details than in free space. **a,** Traditional coded-aperture concept (similar to Figure 2a). **b,** Coded aperture implemented based on spectral diversity of a leaky chaotic cavity (similar to Figure 2e). **c,** Reverberation-coded aperture (RCA) based on spectral diversity: scene and wave sources are inside – as opposed to outside – of the leaky chaotic cavity. **d,** Experimental implementation of a configuration-diverse RCA for a prototypical object-localization task. The panels are reproduced with permission from del Hougne, M. et al., *Phys. Rev. Lett.* **127**, 043903 (2021);[69] licensed under a Creative Commons Attribution (CC BY) license.

The achievable sensing resolution inside an RCA is not determined by the wavelength but can be directly related to the dwell time[209] of waves inside the cavity (before being absorbed) via three distinct RCA mechanisms[69]: (i) enhanced signal strength, (ii) enhanced measurement diversity, and (iii) enhanced intrinsic sensitivity of the field to any perturbation such as the scene[210]. These links between dwell time and achievable resolution were explored in Ref.[69] in the context of a prototypical object localization task (essentially equivalent to imaging an extremely sparse scene), numerically in spectrally-diverse RCAs and experimentally in configurationally-diverse RCAs. A related observation was subsequently also reported in Ref.[211] which numerically studied systems without any absorption: if the average dwell time is constant, so is the average precision with which a particle's property can be estimated, as long as the scattering regime is ballistic or diffusive (but not Anderson localized). The physics of the underlying RCA mechanisms extends of course to more complex tasks, such as the recognition of sub-wavelength object shapes or materials inside an RCA without any manipulation in the extreme proximity of the scene. In this spirit, the results from Ref.[212] can be understood as the successful reconstruction using a spectrally-diverse RCA of random 2D images whose smallest feature size is a sixth of the wavelength; these images were encoded into the configuration of a programmable metasurface in Ref.[212]. In general, besides dwell time and SNR, non-physical parameters like the available *a priori* knowledge and the choice of data analysis algorithm also strongly influence the achievable resolution. Not relying on evanescent waves makes RCAs compatible with the remote sensing requirements of the EM sensing applications we consider – in sharp contrast to the rather invasive examples of computational meta-imagers that reported sub-wavelength resolution by placing meta-hardware in the extreme vicinity of the scene to capture evanescent waves[54,55,58].

Looking forward, an open question is whether self-oscillatory wave fields[213] outperform linear ones for sensing with RCAs. A further challenge is the generation of sufficient training data for the neural network. Because the complicated irregular geometry of a chaotic cavity is unsuitable for simulation, a promising way forward appears to be an electronically reconfigurable test scene to rapidly acquire many examples, as in Ref.[212]. Finally, configuration-diverse RCAs as in Ref.[69] have so far only been used with random coding patterns, and thus pseudo-random scene illumination; one may expect further performance improvements if learned instead of pseudo-random coding patterns are used[176].

**4.3 All-Analog Sensing Pipelines.** We have previously established in section 3 that an intelligent meta-imager performs task-specific analog signal processing "over-the-air" during the measurement process. This naturally leads to the following question: Could the entire sensing pipeline, including the digital part seen in Figure 5a, be implemented in an all-analog fashion? This question is in line with a timely research track on metamaterial wave processors[214] that has emerged completely independently from computational meta-imaging. The underlying idea there is that any wave system can be understood as a mathematical operator acting on the input wavefront, and by suitably tailoring the system one can hope to implement a desired mathematical operation. The promises held by wave-based analog signal processing relate mainly to improving speed and energy efficiency of specific-purpose computing in comparison to the use of a general-purpose digital electronic processor. For example, a simple lens is an entirely passive device that performs a Fourier transform at the speed of light propagation. Speed and energy consumption are obviously also of high relevance to the EM imaging and sensing applications we are addressing.

Our vision of all-analog sensing pipelines hence consists in coupling the intelligent data acquisition discussed in section 3 with analog (as opposed to digital) data processing. In the context of computational meta-imaging, an analog wave processor should satisfy the following criteria to constitute a candidate technology that may replace the current digital layers seen in Figure 5a (typically an artificial neural network, ANN):

1. <u>Programmability:</u> Being able to tune the operation performed by the analog processor is essential for
    (i) switching dynamically between functionalities, e.g., when the meta-imager is to be deployed for a different task,
    (ii) adapting to environmental changes, e.g., to update the ANN in response to changes in the scene (appearance of parasitic scatterers) or measurement hardware (failure of a programmable meta-atom), and
    (iii) transfer learning, e.g., when the ANN is coarsely trained with synthetic data and then fine-tuned *in situ*.
2. <u>Non-Linearity:</u> Non-linear activation functions are a crucial ingredient of ANNs that enables them to tackle highly complex sensing tasks that cannot be solved with linear processing.
3. <u>Integrability:</u> The analog wave processor should be easy to integrate into a realistic

sensing scenario; specifically, it should not require the addition of excessively bulky components.

4. <u>Operation in Native Signal Domain:</u> To avoid slow and energy-inefficient conversions of the captured microwave signals between their native domain and other analog domains (optics, acoustics, electronics, …) or the digital domain, the entire wave processor should operate in the microwave domain without any such conversions.

In the following, we summarize the four leading strategies in the recent context of analog metamaterial wave processors that partially meet the above requirements. On an abstract level, these four different approaches, illustrated in the four rows of Figure 9, all conceive scattering systems that mix the incoming signals in a desired manner.

*4.3.1 Bulk Metamaterial Block.* – The first examples of metamaterial wave processors were carefully designed static metamaterial blocks with a desired linear transfer function[215,216]. While the addition of programmable features to such metamaterial blocks has not been reported to date, the use of materials with intensity-dependent non-linear response has received some attention, albeit only in simulation[217]. Miniaturization to enable integrability has been the key motivation of early works on metamaterial wave processors[215] that had optical implementations as ultimate application in mind. However, experimental devices fabricated for the microwave domain[216], the native signal domain we are interested in for EM sensing applications, remain quite bulky.

*4.3.2 Waveguide Mesh.* – Linear programmable transformations have recently successfully been implemented in guided optics using meshes of programmable Mach-Zehnder interferometers[218–221], building on earlier work that combined free-space propagation with beamsplitters[222]. The addition of non-linear components such as amplifiers to such meshes received attention in the microwave domain very recently[223]. However, such microwave waveguide meshes remain to date quite bulky.

*4.3.3 Cascaded Diffractive Layers.* – Wave propagation through a series of carefully fabricated diffractive layers, often referred to as "diffractive deep neural network", was recently shown to enable the implementation of a desired linear transformation in the terahertz domain[224]. It is important to note that as long as the diffractive layers are linear, adding more layers enhances the ability to implement precisely the desired linear operation[225] but does not add "depth" in the machine-learning sense of the word[175,187] because ultimately the system still performs a linear matrix-vector multiplication[226]. Very recently, Ref.[227] implemented a cascade of programmable linear diffractive layers in the microwave domain. To date, such programmable diffractive layers have not yet been endowed with non-linear features that would be needed for complicated processing tasks. While the compatibility of the concept with CMOS technology for operation at near-infrared frequencies has been demonstrated for static linear systems[228,229], the integrability of cascaded diffractive layers at microwave frequencies appears in general difficult due to their inherent bulkiness and reliance on free-space propagation[227,230].

Most recent works on cascaded diffractive layers can in fact be understood as all-analog sensing pipelines. However, they use task-agnostic instead of intelligent scene illumination, rely on bulky setups and their processing is to date limited to a linear (and in most cases static) projection. Nonetheless, linear processing is sufficient for "simple" classification tasks such as MNIST digit recognition, and ensembles of linear diffractive networks can yield improved performance on "harder" tasks[231]. In most examples, an image is encoded into a wavefront which impinges on the cascaded diffractive layers. The intensities in various sections of the output wavefront are integrated and act as spatial class scores that directly indicate the class of the object displayed in the impinging wavefront. In Ref.[232], the input information is encoded spectrally instead of spatially and sorted into spatial classes. In Ref.[233], spectral instead of spatial object classes are used. Different frequencies illuminate an approximately non-dispersive scene, and the scattered wavefronts interact with the carefully designed dispersive diffractive layers such that the frequency-dependent signal on a single-pixel detector classifies the scene. The intensities in various frequency intervals act as spectral class scores that directly indicate the object class.

***4.3.4 Complex Scattering System.*** – An electrically large irregularly shaped scattering enclosure randomly scrambles incident waves. However, equipped with programmable meta-atoms, its scattering properties can be judiciously tuned such that it performs a programmable linear transformation[234,235]. Ideas for adding non-linearities include the use of diode-loaded ports[236] and input-signal-dependent time-modulation[237]. The bulkiness of 3D volumetric scattering enclosures appears at first sight a major obstacle to integrability. However, on the one hand, integrability can be accomplished by utilizing flat quasi-2D programmable chaotic cavities[73,74]. On the other hand, an alternative approach to achieve integrability is to endow an *existing* enclosure whose primary functionality is completely unrelated (e.g., a metallic toolbox, a microwave oven, an indoor environment, etc.) with a second signal-processing functionality such that no bulky enclosure has to be added to a given operation scenario[234,235]. Instead, any already existing enclosure can be leveraged *in situ* by mounting an ultrathin programmable metasurface on its wall. These ideas enable all-analog processing in the native microwave domain as in Ref.[235], whereas conversions between analog and digital are still an essential part of the proposed schemes in Refs.[234,236,237].

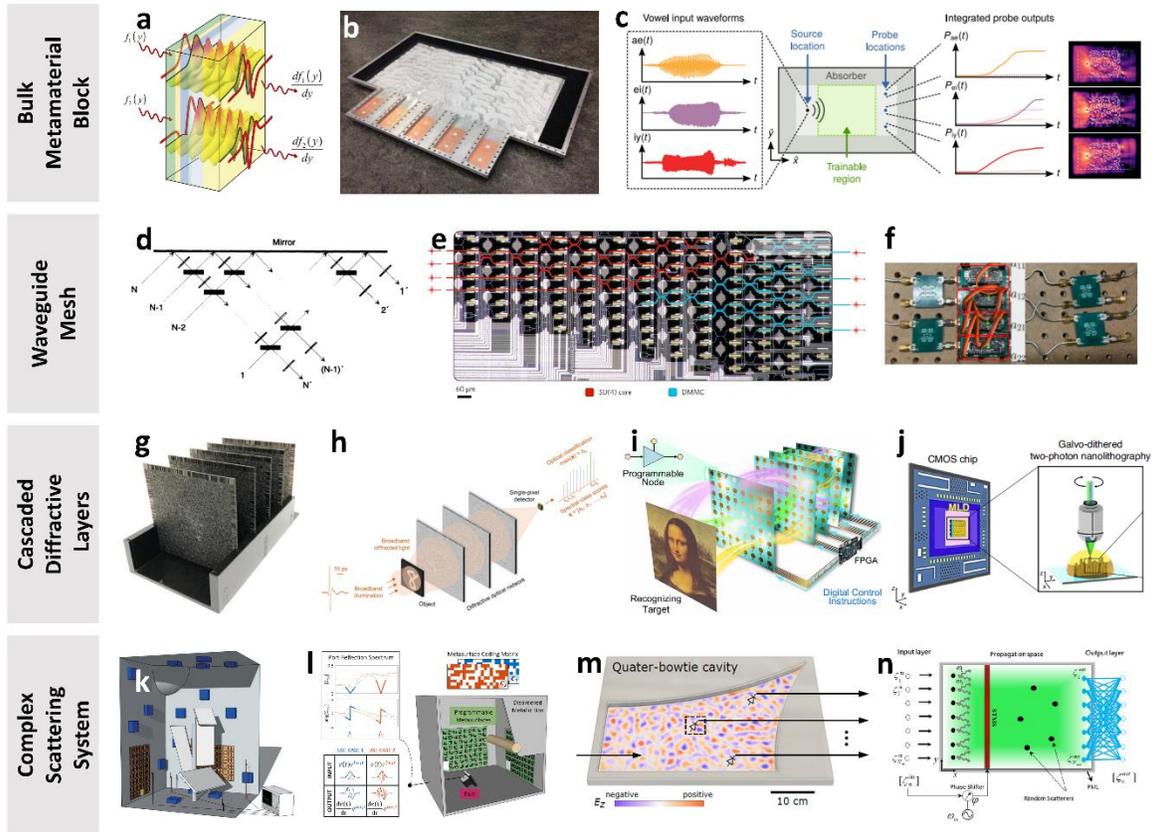

**Figure 9. Metamaterial Wave Processor Paradigms.** First row: bulk metamaterial block (panels adapted with permission from Refs.[215–217]). Second row: waveguide mesh (panels adapted from Refs.[222,221,223]). Third row: cascaded diffractive layers (panels adapted with permission from Refs.[224,227,228,233]). Forth row: complex scattering system (panels adapted with permission from Refs.[234–237]). (Copyrights: **a,** Reproduced with permission from *Science* **343**, 160 (2014).[215] Copyright 2014 American Association for the Advancement of Science. **b,** Reproduced with permission from *Science* **363**, 1333 (2019).[216] Copyright 2019 American Association for the Advancement of Science. **c,** Hughes, T. W. *et al.*, *Sci. Adv.* **5**, eaay6946 (2019);[217] licensed under a Creative Commons Attribution (CC BY) license. **d,** Reproduced with permission from *Phys. Rev. Lett.* **73**, 58 (1994).[222] Copyright 1994 American Physical Society. **e,** Reproduced with permission from *Nat. Photonics* **11**, 441 (2017).[221] Copyright 2017 Springer Nature. **f,** Reproduced with permission from Tzarouchis, D. C. *et al.* Design and implementation of tunable RF modules for reconfigurable metastructures that perform mathematical computations. *in Proc. URSI GASS* (2021).[223] Copyright 2021 Institute of Electrical and Electronics Engineers. **g,** Reproduced with permission from *Science* **361**, 1004 (2018).[224] Copyright 2018 American

Association for the Advancement of Science. **h,** Luo, Y. *et al.*, *Light Sci. Appl.* **8**, 1112 (2019);[233] licensed under a Creative Commons Attribution (CC BY) license. **i,** Liu, C. *et al.*, doi:10.21203/rs.3.rs-90701/v1;[227] licensed under a Creative Commons Attribution (CC BY) license. **j,** Goi, E. *et al.*, *Light Sci. Appl.* **10**, 40 (2021);[228] licensed under a Creative Commons Attribution (CC BY) license. **k,** del Hougne P. *et al.*, *Phys. Rev. X* **8**, 041037 (2018);[234] licensed under a Creative Commons Attribution (CC BY) license. **l,** Reproduced with permission from Sol, J. *et al.*, *arXiv:2108.06178*.[235] Copyright 2021 The Authors (arXiv perpetual non-exclusive license). **m,** Ma, S. *et al.*, doi:10.21203/rs.3.rs-783820/v1;[236] licensed under a Creative Commons Attribution (CC BY) license. **n,** Reproduced with permission from Momeni, A. *et al.*, *arXiv:2107.08564*.[237] Copyright 2021 The Authors (arXiv perpetual non-exclusive license).)

## 5. Conclusions

To summarize, in this Review we retraced the development of computational meta-imagers from the earliest frequency-diverse compressive systems to modern programmable intelligent systems. Compressive meta-imagers (section 2) seek to indiscriminately acquire all scene information with a sensing matrix that represents a near-isometric linear embedding of the scene based on pseudo-random patterns, orthogonal patterns or principal scene components. In contrast, intelligent meta-imagers (section 3) discriminate between task-relevant and task-irrelevant information during the physical measurement process by using learned patterns that are purposefully non-isometric in a way that highlights salient features for the specific task at hand. The acquisition of more task-relevant information per measurement reduces the number of necessary measurements to answer the specific task as well as the processing burden, thereby leading to remarkable improvements in latency.

Intelligent meta-imagers rely on programmable metamaterial hardware. While this hardware has been around since 2014, only recently it was realized that programmable meta-atoms can be integrated as trainable physical weights into a hybrid analog-digital end-to-end sensing pipeline. The latter enables joint training of physical and digital weights, thereby endowing the physical metamaterial layer with task-awareness. Intelligent programmable meta-atoms thereby simultaneously sort and compress scene information, that is, they perform task-specific analog signal processing "over-the-air".

In section 4, we looked forward to the next three major developments that we expect in the field: first, the introduction of cognitive meta-imagers that optimize their settings themselves on-the-fly; second, the integration of the deployment environment's intrinsic scattering and reverberation to achieve remote deeply sub-wavelength resolution; third, the replacement of the currently digital layers with analog metamaterial wave processors that are programmable, non-linear, integrable and operate in the native (microwave) signal domain. Another important avenue for future research relates to the fusion of intelligent meta-imaging concepts with existing wireless communication infrastructure. Thereby, the latter can be endowed with a second sensing functionality. Compared to existing wireless surveillance systems based on WiFi infrastructure[1–5], the available degrees of freedom for advanced sensing will increase by several orders of magnitude through the combination with intelligent meta-hardware. Thereby, the required sensing tasks for situational awareness can be performed by recycling existing EM waves and transceivers, avoiding the hardware and spectrum allotment cost of antenna networks dedicated to sensing[6–14]. A further opportunity lies in the integration of information from non-EM sensors (optical, mechanical, …) into the intelligent sensing pipeline. Current intelligent meta-imagers target EM sensing and monitoring applications that will enable remote situational awareness, which is highly demanded in security, health care, mobility, and human-machine interaction. We also faithfully expect that many meta-imagers based on other wave phenomena (acoustic, elastic, terahertz, …) can be endowed with programmability where not yet available (e.g., Refs.[238,239] for acoustics) and subsequently upgraded to the modern intelligent meta-imager paradigm that we have reviewed.

## Acknowledgements

The authors thank their international collaborators on various aspects of the presented review for stimulating discussions.

## Additional Information

**Competing Interests**: The authors declare no competing interests.

**Data Availability**: Data sharing is not applicable to this article as no new data were created or analyzed in this study.